\documentclass{article}
\usepackage[utf8]{inputenc}
\usepackage{times}
\usepackage{graphicx}
\usepackage{amssymb}
\usepackage{amsthm}
\usepackage{amsmath}
\usepackage{dsfont}
\usepackage{bm}
\usepackage{mathrsfs}
\usepackage{bbold}
\usepackage{gensymb}
\usepackage{doi}
\usepackage{soul}
\usepackage{ulem}
\usepackage{authblk}

\title{The next generation of analogue gravity experiments}
\author[1]{M. J. Jacquet}
\author[2]{S. Weinfurtner}
\author[3]{F. K\"onig}
\affil[1]{ Laboratoire Kastler Brossel, Sorbonne Universit\'{e}, CNRS, ENS-Universit\'{e} PSL, Coll\`{e}ge de France, Paris 75005, France\\maxime.jacquet@lkb.upmc.fr}
\affil[2]{School of Mathematical Sciences, University of Nottingham,University Park, Nottingham, NG7 2RD, UK} 
\affil[3]{School of Physics and Astronomy, SUPA, University of St. Andrews, North Haugh, St. Andrews, KY16 9SS, United Kingdom}

\begin{document}

\maketitle

This article is an introduction for a theme issue following a Scientific Discussion Meeting on \emph{The next generation of analogue gravity experiments} held at the Royal Society in December 2019. This theme issue comprises a collection of recent advances of the research programme, as well as their philosophical implications, that were presented at the meeting.

Analogue gravity~\cite{barceloAnalogueGravity2011} summarises an effort to mimic physical processes that occur in the interplay between general relativity and field theory in a controlled laboratory environment. The aim is to provide insights in phenomena that would otherwise elude observation: when gravitational interactions are strong, when quantum effects are important, and/or on length scales that stretch far beyond the observable Universe. The most promising analogue gravity systems up-to-date are fluids, superfluids, superconducting circuits, ultra-cold atoms and optical systems. While deepening our understanding of the laboratory systems at hand, the long term vision of analogue gravity studies is to advance fundamental physics through interdisciplinary research, by establishing and nurturing a new culture of collaboration between the various communities involved.

The common feature all analogue gravity systems share, is that small excitations within exhibit dynamical equations that can be mapped to the equations one usually encounters in classical and quantum field theory in curved spacetimes~\cite{unruh_experimental_1981}. The excitations experience an effective spacetime geometry, that is completely determined by the propagation speed of the excitations and their relative propagation speed with respect to their medium. Control over these quantities enables the engineering of a variety of d+1~\footnote{Analogue gravity uses the standard notation from general relativity to express the dimensionality of spacetime: d is the number of spatial dimensions, and $1$ denotes one temporal dimension.} dimensional spacetime geometries~\cite{visser_acoustic_1998}.

While the overall idea of analogue gravity experiments has been around for 40 years, the first experimental demonstrations and successes are fairly recent. The evolution of analogue gravity studies can be divided into three phases: the `early years' (between 1981-2008) when analogue gravity was mainly a theoretical field of research, followed by an 'experimental age' with an explosion of analogue gravity studies focussing mainly on the observation of the Hawking effect in a variety of media between 2008-2017~\footnote{Two notable exceptions were studies of the dynamical Casimir effect in superconducting circuits~\cite{Wilson_Casimir_2011} and in atomic Bose-Einstein condensates~\cite{Westbrook_Casimir_2012}.}. Over the last few years a new wave of experiments appeared, facing different challenges attempting to extend the analogy beyond the observation of the Hawking effect in the laboratory. In this spirit, a new 'experimental age' has started whose full potential has not yet been reached. A main focus of the Meeting was to discuss and shape the future, or the next generation of analogue gravity experiments.

Let us briefly review the history of the field and flag up important results established over the last decade to set the stage for this issue. The reader will also find a focused literature review at the beginning of each article.

%
\section{A brief history of analogue gravity}
%
The beginning of analogue gravity dates back to Unruh's seminal paper in 1981~\cite{unruh_experimental_1981}, where he demonstrates the mathematical analogy between the dynamics of sound waves in a supercritical fluid flow and fields at the event horizon of gravitational black holes. By doing so, he was able to raise the possibility of studying the black hole evaporation or Hawking effect in a controlled laboratory setting~ \cite{hawking_black_1974,hawking_particle_1975}.
At the turn of the 21st century, analogue gravity theory studies boomed, and it was shown that the gravitational analogy can be extended to a broad range of physical systems~\cite{novello_artificial_2002,volovik_universe_2003,barceloAnalogueGravity2011,faccio_phenomenology_2013}.

Despite the early theoretical developments, it took until 2008 for experiments to follow. These pioneering experiments involved the scattering of effective one dimensional horizons in two very different media: surface waves in an open channel flow~\cite{rousseaux_observation_2008} and light propagation in an optical fibre~\cite{philbin_fiber-optical_2008}. Two years later, experimental setups using bulk crystals~\cite{faccio_analogue_2010} and Bose-Einstein condensates~\cite{lahav_realization_2010} were readily available to set up effective 1+1 dimensional black hole geometries. In 2011, building up on~\cite{rousseaux_observation_2008}, the first excitation spectrum for surface waves in an open channel flow from an effective white whole horizon was observed~\cite{weinfurtner_measurement_2011}. The observed spectrum was in agreement with the Hawking effect. 
Within this short period of three years, through these experiments analogue gravity was transformed from a theory-driven to an experiment-driven line of research.

Over the course of the next decade, the experimental realisation of effective black hole horizons made notable progress. The tunnelling of light waves across an optical horizon was observed in 2012~\cite{choudhary_efficient_2012}. An important ingredient of the Hawking effect is the generation of negative frequency waves, as for example observed in the open-channel flow experiments~\cite{weinfurtner_measurement_2011}. A series of experiments between 2012 and 2018 demonstrated the generation of negative frequency waves in optical fibres and bulk systems~\cite{rubino_negative-frequency_2012,mclenaghan_compression_2014,Jacquet_book_2018}. 

In 2015 a new analogue gravity platform emerged: fluids of light realised with polaritons in semiconductor microcavities, in which sonic horizons were demonstrated~\cite{nguyen_acoustic_2015}. In the following year the open channel-flow experiments made significant advances by extracting Hawking correlations from classical noise on the air-water interface surrounding an effective white hole horizon~\cite{Rousseaux_PRL_2016} leading to~\cite{euve_scattering_2020}. In the same year, a first study on the extraction of Hawking correlations from an effective black hole horizon in an atomic Bose-Einstein condensate appeared~\cite{steinhauer_hawking_2016}. Over the years, an increase in the sensitivity and control over the setup enabled the extraction of the Hawking spectrum~\cite{munoz_de_nova_observation_2019} and the exploration of the Hawking spectrum from an evolving effective black hole horizon~\cite{kolobovSpontaneousHawkingRadiation2019}. In the same period, the stimulated emission of Hawking paired waves (including the negative frequency partner) was observed from an optical fibre system~\cite{drori_observation_2019}. Last but not least, in 2019, the first realisation of an effective white hole horizon for sound waves in superfulid $^{3}\mathrm{He}\text{\ensuremath{-}}B$ was achieved~\cite{Skyba_whitehole_2019}.

In parallel to the ongoing effort on observing the Hawking effect in the laboratory, a new line of investigation is concerned with 2+1 dimensional analogue black holes. The additional spatial dimension enables the investigation of  scattering processes around rotating black holes. A study of surface-waves scattering off a water vortex revealed black hole superradiance~\cite{torres_rotational_2017}, and a subsequent study of the noise on the free surface surrounding the vortex flow led to the observation of black hole ringdown modes~\cite{torres_lightrings_2018}. A rotating black hole was also demonstrated in a fluid of light in a thermo-optical defocusing medium~\cite{Vocke_rotating_2018}. Rotational superradiance may also occur from rotating objects, as exemplified by the amplification of twisted sound waves propagating through rapidly rotating disks~\cite{Faccio_zeldovich_2020}. An intriguing new direction of research within rotating spacetime geometries is to employ topological defects (pairs of vortices and anti-vortices) representing particles in an effective relativistic setting to extract energy from a central vortex in a Bose-Einstein condensate by the Penrose effect \cite{Solnyshkov_penrose_2019}, the particle equivalent of superradiant wave-scattering. 

Alongside the above development on reproducing black hole physics, experiments mimicking the Universe as a whole have emerged. Effective black hole horizons are set up by means of a spatially varying background, while cosmological scenarios are realised via a rapid temporal change in the background parameters. The first analogue experiments of this kind were investigations of the dynamical Casimir effect in a superconducting circuit~\cite{Wilson_Casimir_2011} and in a parametrically excited Bose-Einstein condensate~\cite{Westbrook_Casimir_2012}. More recently, two new experimental studies have appeared: in 2018 sound waves in a rapidly expanding ring-shaped Bose-Einstein condensate were shown to exhibit cosmological redshift and Hubble friction~\cite{eckel_expanding_2018}. The year after, a controlled, rapid change in the trapping parameters of an ion-chain resulted in the creation of phonon pairs akin to cosmological pair creation \cite{wittemerPhononPairCreation2019a}. A promising direction of experimental analogue gravity studies within cosmological scenarios was presented at the meeting: a coupled two-component Bose-Einstein condensate can be used to mimic a first order relativistic phase-transition (also known as the false vacuum decay) ~\cite{bradenNewSemiclassicalPicture2019}. This is akin to an evolving vacuum state of the Universe.

This brief summary of experimental progress is far from being exhaustive. Our selection displays the consistently growing number of communities getting involved in analogue gravity studies, as well as the growing numbers of fundamental physics processes under investigation.

\section{Summary of the issue}
We now focus our attention on the content of this theme issue on \emph{The next generation of analogue gravity experiments}. This comprises 11 articles, which touch on recent developments at the forefront of theoretical and experimental analogue gravity studies in both effectively 1+1 or 2+1 dimensional spacetime geometries. They also fairly represent the diversity of media in which one can study field theories on curved spacetimes. 

The order of articles loosely follows that of the sessions of the Scientific Discussion Meeting. Each article may be read independently from the others. However, we encourage readers who are not familiar with analogue gravity to start with the first article, entitled ``Hawking radiation in optics
and beyond'' by \textbf{Raul Aguero-Santacruz and David Bermudez}, as this also provides an introduction to analogue gravity more generally. The authors review the fundamentals of the physics of the Hawking effect from black hole horizons.
They then use the analogue gravity setup to re-derive Hawking's result and thus introduce the essential mathematical tools of quantum field theory on curved spacetimes.
Having shown how quantum fluctuations at the horizon yield spontaneous emission in entangled pairs, they explain how this could be observed in an optical setup.
Finally, they show how seeding the Hawking effect with a classical input state stimulates emission.
In line with current philosophical debates, the authors conclude their article with considerations touching the epistemology of analogue gravity.
\textbf{Jack Petty and Friedrich K\"onig} also use the optical platform to investigate the amplification of coherent, classical fields.
Specifically, they contrast amplification at the horizon with resonant radiation (also known as optical \v{C}erenkov radiation).
They discuss the role dispersion plays in the kinematics and dynamics of both processes and discover a regime of record efficiency of amplification of resonant radiation. 
This highly tuneable laser source provides a novel application of analogue gravity physics.
\textbf{Yuval Rosenberg} revisits the landmark ideas and experiments for optical systems.
Throughout his article, he demonstrates how dispersion enables the physics at play: he shows how it is essential to the formation of horizons, the generation of negative frequency waves and spontaneous as well as stimulated emission by the Hawking effect. 
He also reviews explanations of the analogue gravity system specific to optics.
He concludes his paper by calling for analogue gravity optical experiments ``beyond the horizon''.
\textbf{Ulf Leonhardt} draws inspiration from the use of transformation optics in analogue gravity, to guide us through considerations on cosmology. He presents the Lifschitz theory of the cosmological constant that he has recently developed, and argues that future experiments of analogue gravity could test these predictions.

\textbf{Tobias Sch\"atz} and collaborators look back upon their recent observation of cosmological particle pair-creation and show how machine-learning strategies can be used to increase experimental control on the motion of a single ion, for example.
This opens an avenue for the future generation of spatial entanglement amongst pairs of ions, which they propose to characterise by a measure of squeezing.
Such experimental techniques could be used to investigate other effects, such as the Hawking effect, or even go beyond analogue gravity to investigate for example the Sauter-Schwinger effect.
In a similar spirit, \textbf{Elisabeth Giacobino} and collaborators propose to characterise the output state in their experiments by means of squeezing and entanglement. In their article, they look back on polariton flows in semiconductor microcavities and demonstrate how various flow profiles in 1+1 and 2+1 dimensions may be optically engineered.
The versatility of their method is exemplified for non-rotating and rotating geometries.
Therefore, the authors argue that fluids of light, such as polaritons, are an ideal platform to measure \textit{e.g.} the Hawking effect.
\textbf{Miles Blencowe and Hui Wang} are also interested in devising experiments to measure quantum properties in analogue gravity experiments, here with focus on both, the Hawking and Unruh effects.
Notably, they bring superconducting circuits back to the foreground by re-deriving the Hawking temperature in the light of new experimental possibilities offered by newly developed Josephson travelling-wave parametric amplifiers.
They also discuss an ``oscillating'' scheme to generate Unruh radiation and use the logarithmic negativity as an entanglement measure to theoretically show that entangled pairs are indeed emitted in their proposed setup.

\textbf{Germain Rousseaux and Hamid Kellay} share with us the ``plumber's expertise'': they summarise the necessary tools to observe the Hawking effect in 1+1 dimensional water experiments and insist on the combined influence of hydrodynamics and dispersion on the output spectrum.
In the second part of their article, they enrich the family of analogue gravity experiments with a new experimental platform, flowing films of soap, and demonstrate the creation of flows with horizons.

Going from uni-dimensional motion to rotating geometries, \textbf{Theo Torres} calculates the spectrum of superradiance in dispersive media.
His careful analysis of the system reveals that some waves are partially reflected by the drain that generates the vortex flow.
This observation is in contrast to the behaviour of waves on Kerr black holes.
And yet, the interplay between vorticity and dispersion does not prevent superradiant amplification of incoming waves at the ergosurface.
Torres builds on this demonstration of the robustness of the effect to encourage researchers to push their platforms beyond the strict ``analogue regime'' in search of new effects of waves in media.
\textbf{Cisco Gooding} also writes about superradiance: he considers the case of acoustic superradiance --- the amplification of sound waves on a rotating, absorbing cylinder.
He shows in which regimes sound waves with orbital angular momentum may be amplified, and demonstrates once more that dispersion must be included in all calculations aiming at providing realistic predictions for experiments.

In the last article of this theme issue, \textbf{Karim Thebault and Peter Evans} look back on the analogue gravity programme from a philosopher's perspective.
They ask again ``what can be learnt'' from analogue gravity experiments and thus question notions of universality and validation, among others.
They claim that neither the accessibility nor the manipulability of astrophysical black holes (and their event horizons wherefrom Hawking radiation originates) are necessary to obtain experimental knowledge about the Hawking effect.
This original claim rests on the use of inductive triangulation to set the limits of experimental knowledge.
Their work highlights the tension on when exactly reasonable doubt has been mitigated --- a matter that reaches far beyond analogue gravity.

On this note, we would like to conclude this introduction and line of thoughts as follows: the limits of our experimental knowledge are only ever contextual, with time we may always push them further.
This is what the next generation of analogue gravity experiments shall aim at.

\subsection*{Acknowledgements}We thank the Royal Society Hooke Committee for giving us the opportunity to organise the Scientific Discussion Meeting from which this theme issue stems, and Annabel Sturgess and her team for hosting us at the Royal Society. We are also thankful to Alice Power for editing this issue with MJJ. We thank the authors of articles of the issue for their feedback on this manuscript. Lastly, we wholeheartedly thank all the participants of the meeting for their contribution to the ongoing discussion on analogue gravity.

\end{document}